# Tunable Lifshitz Transitions and Multiband Transport in Tetralayer Graphene

Yanmeng Shi[1], Shi Che[1,2], Kuan Zhou[1], Supeng Ge[1], Ziqi Pi[1], Timothy Espiritu[1], Takashi Taniguchi[4], Kenji Watanabe[4], Yafis Barlas[3], Roger Lake[3], and Chun Ning Lau[1,2*]

[1]Department of Physics and Astronomy, University of California, Riverside, Riverside, CA 92521, USA
[2] Department of Physics, The Ohio State University, Columbus, OH 43210, USA
[3]Department of Electrical and Computer Engineering, University of California, Riverside, Riverside, CA 92521, USA
[4]National Institute for Material Science, 1-1 Namiki, Tsukuba, Ibaraki 305-0044, Japan

**As the Fermi level and band structure of two-dimensional materials are readily tunable, they constitute an ideal platform for exploring Lifshitz transition, a change in the topology of a material's Fermi surface. Using tetralayer graphene that host two intersecting massive Dirac bands, we demonstrate multiple Lifshitz transitions and mutiband transport, which manifest as non-monotonic dependence of conductivity on charge density *n* and out-of-plane electric field *D*, anomalous quantum Hall sequences and Landau level crossings that evolve with *n, D* and *B*.**

The Fermi surface of metals and semiconductors can undergo Lifshitz transition (LT), *i.e.* abrupt topological changes in the connectivity of the band structures, giving rise to anomalies in the material's properties [1]. For instance, it is associated with the onset of superconductivity in iron-based superconductors[2-4] and plays an important role in the properties of Weyl semimetals[5,6]. In bulk materials, as the Fermi energy is not easily tuned [7], such a transition and the accompanying anomalies are difficult to observe, though transitions as a function of magnetic field or temperature has recently been observed in heavy fermion compounds [2,8] and bulk WTe$_2$ [9]. In principle, this difficulty is bypassed with the recent advent of two-dimensional (2D) materials, which allows *in situ* tuning of the Fermi level and even band structure by external gates [10-12], though observation of LT remains scarce [13,14] due to the stringent requirement of high mobility. Direct evidence of LT, particularly in the absence of external fields, is limited, as the relatively low energy at which LT occurs (~ 1 meV) can be easily obscured by disorder or symmetry-breaking electronic interactions.

Here we focus on a relatively unexplored 2D multi-Dirac band system, Bernal-stacked tetralayer graphene (4LG) [15-17]. Its band structure can be decomposed into two intersecting bilayer graphene (BLG)-like bands with light and heavy effective masses, which are hybridized at low energies due to next-nearest-layer hopping [17]. The hopping term $\gamma_3$ between skewed lattice sites in adjacent layers gives rise to trigonal warping that persists to relatively large energies, ~ 10 meV. The co-presence of both bands, combined with trigonal warping, provides a tunable platform in which LT is readily accessible. Using high quality 4LG devices, we observe dramatic features in transport characteristics, including non-monotonic dependence of conductivity on charge density *n* and out-of-plane electric field *D*, anomalous quantum Hall (QH) sequences and Landau level (LL) crossings. These features are identified to arise from as many as six different changes in topology of Fermi contours and multiband transport as *n, D and B* are varied. The large number of LTs, and the number of *in situ* variables that drives them, underscore the rich physics in 4LG that, in contrast to most material systems affording LT, constitutes of only a single element.

---

[*] Email: lau.232@osu.edu

Hexagonal boron nitride (hBN)-encapsulated 4LG devices are fabricated using a dry transfer technique [18,19] and measured in cryostats at low temperature (~ 0.26 - 1.5K) using standard lock-in techniques. With the doped silicon back gate and the metal top gate, we are able to tune charge density *n* and perpendicular electric field *D* independently, where $n = (C_{tg}V_{tg} + C_{bg}V_{bg})/e$ and $D = (C_{tg}V_{tg} - C_{bg}V_{bg})/2\varepsilon_0$, $C_{tg}$ ($C_{bg}$) is the geometry capacitance per area between graphene and the top (back) gate, $V_{tg}$ ($V_{bg}$) the gate voltage applied to top (back) gate, *e* the charge of electron, and $\varepsilon_0$ the permittivity of vacuum.

Fig. 1a displays the longitudinal resistance *R* of 4LG as a function of *n* and *D* in the absence of magnetic field. At the charge neutrality point (CNP), *R* firstly increases symmetrically with *D*, indicating the opening of a band gap due to the broken inversion symmetry, then saturates at larger *D* > ~ ± 220 mV/nm (Fig. 1b). This is consistent with theoretically predicted opening of a small band gap that saturates at ~ 5 meV for large *D* [20].

A close examination of Fig. 1a reveals several rather surprising features. In particular, what immediately sets it apart from thinner graphene devices is that, instead of a single peak at the CNP, *R(n)* of 4LG exhibits *three* peaks, which are labeled as *X, Y* and *Z*, respectively (Fig. 1c). The leftmost peak *X* is the most prominent, and rightmost peak *Z* is present as a small shoulder. Such non-monotonic *R(n)* behavior points to the underlying multibands of tetralayer graphene that are more complex than its thinner graphene counterparts [21-27]. In fact, the charge neutrality point (CNP), as determined from the peak in *R* at large *D*, is *not* located at any of the three peaks at *D*=0. Instead, it corresponds to a local resistance *minimum* located between peaks *X* and *Y* (Fig. 1c). This identification of the CNP with a local resistance minimum represents yet another "deviation" from the standard behaviors of mono-, bi- and tri-layer graphene devices, where the CNP is invariably associated with a resistance peak [21-23,25,27-29]. Another unexpected feature is the intricate dependence of the peaks on *D* – as |*D*| increases, peak *X* appears to move to larger charge density, while peak *Z* splits into two peaks, $Z_1$ that is relatively stationary in *D*, and $Z_2$ that moves linearly with *D*. The movement of the peaks *X* and $Z_2$ can be clearly seen in line traces *R(n)* at *D* = -125 and -200 mV/nm (Fig. 1d).

To understand these unusual features in the *R(n, D)* map, we calculate the band structure of 4LG using the tight-binding model [17]. The hopping parameters are extracted by matching the LL crossing points between experimental data *R(n, B)* over a wide range of *B* at *D*=0 and LL spectra that are calculated using a *k.p* continuum model [16,17], similar to procedures that are used to determine hopping parameters in trilayer graphene [26,30,31] (for details, see Supplemental Material [19]). Fig. 2a-b display the three-dimensional (3D) band structure and its 2D projection $E(k_x)$ at *D*=0, respectively. At higher energies, the bands are well approximated by two intersecting BLG-like bands with different effective masses. The band structure calculated by ignoring the off-diagonal blocks in the Hamiltonian are plotted as the dotted lines in Fig. 2b, showing the outlines of the intersecting bands. Roughly speaking, the effective masses are related to that of stand-alone BLG by the golden ratio [17].

Using the hopping parameters obtained from fitting, we estimate $m^* \sim 0.05\ m_e$ for the heavy mass band, and $0.025\ m_e$ ($0.031\ m_e$) for the conduction (valence) band of light mass band (here $m_e$ is the rest mass of electrons).

At low energies, *next-nearest-neighbor* hoppings hybridize the bands (solid lines, Fig. 2b). In particular, as a result of the skewed hopping term $\gamma_3$, trigonal warping significantly distorts the band structure [11-13,32] – as we tune the Fermi level $E_F$ of 4LG by gating, the topology of the Fermi surfaces changes, leading to anomalies in the resistance. A series of cross-sections of the band structure showing the Fermi surfaces evolution is plotted in Fig. 2c. For instance, at the CNP, 4LG is a semi-metal with overlapping conduction and valence bands (Fig. 2b inset); its Fermi surface consists of three elongated pockets of holes and three circular pockets of electrons, which occupy the edges and vertices of a triangle, respectively. These pockets of electrons and holes are isolated from one other. As more holes enter the device, the electron pockets disappear while the three disjoint hole pockets expand. When $E_F$ is lowered past the point at which the light mass and heavy mass bands intersect, $E = -10$ meV, the Fermi surface morphs from a doubly connected triangle with a hole in the center to a singly connected triangle. This abrupt change in the topology of the Fermi surface is reflected in the $R(n)$ data, and is associated with the resistance peak $X$ in Fig. 1. Similarly, when $E_F$ increases from the CNP to $\sim 4.6$ meV, the resistance peak $Y$ accompanies the LT as the topology of the Fermi surface evolves from three disjoint pockets of electrons to a doubly connected triangle; with further increase of $E_F$ to 10 meV, the Fermi surface evolves into a singly connected triangle with the presence of both BLG-like bands, giving rise to the resistance peak $Z$. These energies are labeled $\varepsilon_X$, $\varepsilon_Y$ and $\varepsilon_Z$ in Fig. 2b, respectively. In other words, each of the three peaks in resistance, as well as the minimum at the CNP, are associated with distinct LTs as the Fermi energy varies.

Upon the application of an interlayer potential difference $\Delta$, 4LG's inversion symmetry is broken, leading to non-trivial modification of the band structure. For instance, for $\Delta > 20$ mV, a small band gap opens, giving rise to the observed high resistance at the CNP. Fig. 2d-e plot the 3D band structure and 2D $E(k_x)$ at $\Delta = 25$ mV, respectively, while Fig. 2f plots the evolution of Fermi surfaces at $\Delta = 25$ mV. Note that $\Delta$ denotes the actual potential bias across the 4LG; because of screening, it is typically much smaller than that imposed by external gates, $Dd$, where $d \sim 1$ nm is the separation between the outmost layers. The Lifshitz transition persists at finite $\Delta$ -- when $E_F$ moves away from conduction and valence band edges, the topology of the Fermi surface changes from three disjoint pockets to a hollow triangle, to finally a singly connected triangle.

Another important effect of $\Delta$ is the lifting of the accidental degeneracies at $\varepsilon_Z$ and $\varepsilon_X$, as it causes the light-mass bands to split away from zero energy. In other words, the two intersecting BLG-like bands at $\Delta=0$ split into two sub-bands, with energetic separations that approximately scale with $\Delta$. Alignment of the Fermi level with the edges of the second sub-bands, which host electrons with very low velocities, provide additional channels for scattering, thus leading to the resistance peaks at the corresponding charge densities, *i.e.* this degeneracy breaking by $\Delta$ gives rise to the split peaks $Z_1$ and $Z_2$ in the electron-doped regime, and $X$ in the hole-doped regime. Quantitatively, we use the Boltzmann transport theory to calculate 4LG's resistivity as a function of $n$ and interlayer potential $\Delta$, as shown in Fig. 1e. The simulation satisfactorily reproduces the major features in the data. To account for the local minimum resistance at the CNP, we take the inter-band scattering between conduction and valence bands into account, which produces a local resistivity minimum near $n = D = 0$, suggesting that such inter-band scattering may play an important role in the transport of 4LG, particularly at the CNP.

Evidence for Lifshitz transition and multiband transport is also visible in the magnetotransport data. In a sufficiently high magnetic field, the cyclotron orbits of electrons coalesce into Landau levels, which reflect the local topology of the Fermi surface. Fig. 3a displays the low field magnetotransport data $R(n, B)$ at $D = 0$ for $0 < B < 3$ T. The filling factors of single-particle QH states jump by 4 as a result of four-fold (valley and spin) degeneracy of each LL, as indicated by the numbers. At higher fields, QH states at filling factors $\nu$=8, 12, 16... are well-resolved. The two sets of LL with different slopes in $B$ arise from the two BLG-like bands, leading to a rich pattern of Landau level crossings.

To account for the magnetotransport data, we calculate the LL spectrum at $D = 0$, as shown in Fig. 3b. In the electron-doped regime, the two sets of LLs converge to $E \sim -5$ and $E \sim 10$ meV, respectively, corresponding to the heavy-mass and light-mass bands. Notably, at very low energies, owing to the trigonal warping of the Fermi surface into three disjoint "pockets", the lowest few LLs are "bundled" into groups of three, Such bundling is in addition to the spin and valley degeneracy, thus the LLs are 12-fold degenerate at low energies. This triplet is broken at slightly higher $B$, and the three levels disperse with $B$ in a complicated pattern, resulting in a number of level crossings. Thus, we expect dual manifestation of trigonal warping and LT in the low-field magnetotransport data: (1). QH states at filling factors $\nu = 12$ and 24 that are resolved prior to that at $\nu = 8$ (one might naively expect the latter from the band structure at higher energies), and (2). LL crossings that emerge at very low energy, *i.e.* below 1.5T and close to the CNP. Both features are borne out in experimental data – as shown in the high resolution plot of $dR/dB$, the $\nu$=12 and 24 QH states are indeed resolved at $B<\sim0.3$T, and a number of LL crossing points are visible near $n = 0$, as indicated by the dotted square (Fig. 3c).

For a direct comparison to data, we also compute the total density of states, which is approximately proportional to the longitudinal resistance, as a function of $n$ and $B$ (Fig. 3d). Overall the agreement with experimental data is excellent; in particular, all the LL crossing points are well-reproduced. We note that the simulation and the data differ at the CNP – the former displays small DOS while the data exhibits large $R_{xx}$. Such a discrepancy arises from electronic interactions that may induce symmetry-broken states at dilute doping, similar to that observed in bilayer [28,33-35] and trilayer [29] graphene. Such interaction-induced effects are not accounted for by our single-particle simulation, and will be studied in a future work.

Lastly, we examine the longitudinal resistance at constant $B$ as $n$ and $D$ are modulated. The $R_{xx}(\nu, D)$ at $B = 2$ T presents a striking pattern (Fig. 4a), where the blue regions represents the gapped incompressible states at integer filling factors, separated by the bright compressible states. Superposed on the background of blue vertical stripes are the series of bright, interrupted lines that diverge from equally spaced spots at $D$=0, indicated by the yellow stars, and visually resemble an interference "ripple" pattern. To quantitatively account for the observed patterns, we calculate the LL spectrum as a function of $D$ at $B = 2$ T (Fig. 4b), where red solid and blue dotted lines represent K and K' valleys, respectively. Notably, the evolution of a given LL from a single valley is markedly asymmetric in $D$, though the combined spectra from both valleys is always symmetric, as expected from the condition $E(D, K) = E(-D, K')$ imposed by the lattice's inversion symmetry. The effect of trigonal warping is evident as the robust $\nu = -12$ state. The "ripple"-like pattern emerges from the intersecting sets of LLs that originate from the light mass and heavy mass bands, respectively. For instance, at $D = 0$, the lowest electron-like LLs of light mass band (b, 0) and (b, -1) reside at $\varepsilon_Z$, the accidental degeneracy point of the BLG-like bands, and crosses the $N = 2$ LL of the heavy mass band (B, 2) at $B=2T$. As the effective masses of the two bands differ by a factor of ~2, the crossings repeat for higher LLs, leading to the equally spaced bright spots at $D = 0$. Application of $D$ lifts this accidental degeneracy as well as the valley degeneracy. Energies of the (b, K') LLs increase rapidly with $D$, whereas those of (b, K) decrease weakly with $D$, giving rise to the LL crossing points that move to larger $\pm D$ at higher $n$. Thus the interference-like pattern is a direct manifestation of the multi-band magnetotransport that is tunable by $B$ and $D$. The simulated density of states vs. $B$ and $D$ is shown in Fig. 4c, and reproduces the experimental data very well. Furthermore, by matching the crossing points in the ($v$, $D$) plane, we are also able to estimate that $D$ as imposed by external gates is screened by a factor of 5.6. In other words, the effective dielectric constant of 4LG is 5.6, assuming the linear relationship.

Bernal-stacked 4LG is a highly tunable system. The trigonal warping induced by the skewed hopping term, combined with the presence of two bands with different effective masses, gives rise to multiple Lifthitz transitions that manifest as resistance peaks or dips at $B = 0$; these transitions can be tuned by a perpendicular electric field and charge density. In the QH regime, the Lifshtiz transition and the two-band transport are evident as the robust states at $v = 12$ and 24 at very low fields, and intricate LL crossing patterns that depend on $n$, $B$ and $D$. What distinguishes this from LT previously observed in BLG [13] is the number of transitions revealed by transport data, its visibility in the absence of external magnetic and electric fields, its tunability by both $n$ and $D$, and its persistence to relatively high energy, ~ ±10 meV, which is an order of magnitude larger than that in BLG. As an electronic system with single element, 4LG provides a simple yet rich platform to explore the topology of Fermi surface and its effect on material properties.

The experiments are supported by DOE BES Division under grant no. ER 46940-DE-SC0010597. The theoretical works and collaboration between theory and experiment is enabled by SHINES, an Energy Frontier Research Center funded by DOE BES under Award # SC0012670. Growth of hexagonal boron nitride crystals was supported by the Elemental Strategy Initiative conducted by the MEXT, Japan and a Grant-in-Aid for Scientific Research on Innovative Areas "Science of Atomic Layers" from JSPS.

FIG. 1. (a) Longitudinal resistance map $R(n, D)$ in log scale at $B=0$. The unit is k$\Omega$. (b) Line trace $R(D)$ at the CNP. (c-d) Line traces $R(n)$ at $D = 0$ (c), -200 (d, red) and -125 (d, blue) mV/nm, respectively. The traces are offset by 0.2 k$\Omega$ for clarity in (d). The black arrows indicate the peaks and the CNP. (e) Differentiated simulated d$R$/d$n$($n$, $D$) using Boltzman transport theory at the same charge density as (a).

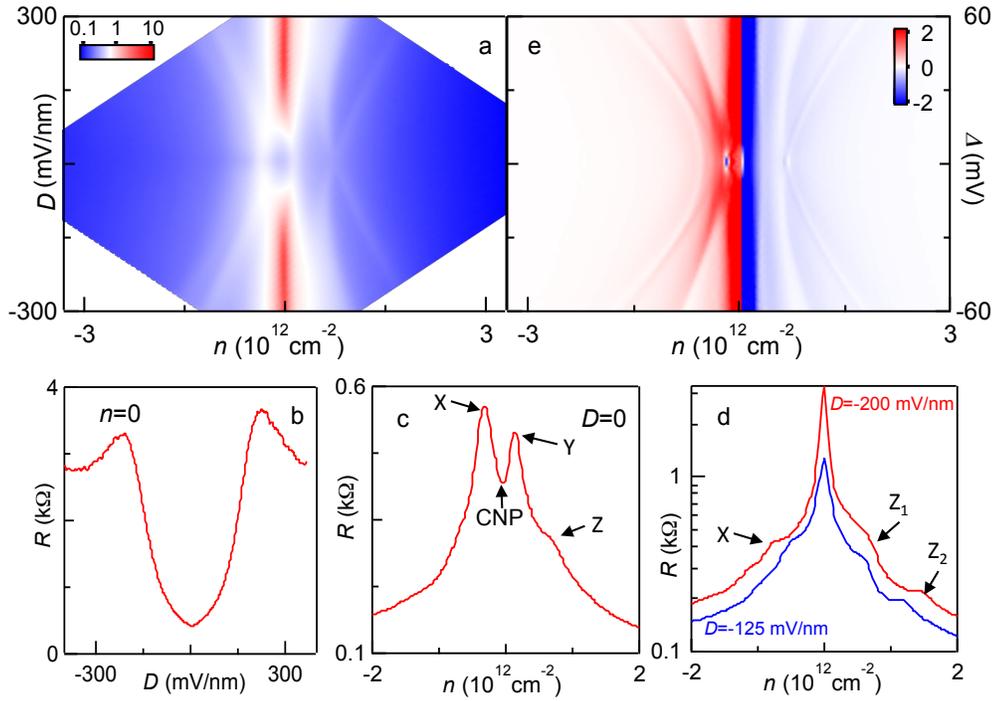

FIG. 2. (a - b) Calculated 3D band structure and its 2D projection $E(k_x)$ at $\Delta = 0$, respectively. In (b), the inset displays the overlap of conduction and valance bands at the CNP. Horizontal dashed lines indicate the energy levels corresponding to the Lifshitz transition points. The red solid and dotted lines are calculated by taking the full Hamiltonian and neglecting the off-diagonal block in Hamiltonian, respectively. (c) A series of cross-sections of band structure showing evolution of the Fermi surfaces by tuning Fermi energy at $\Delta = 0$. (d-e) Calculated 3D band structure and 2D projection $E(k_x)$ at $\Delta = 25$mV. A band gap at the CNP is opened, and the accidental degeneracy at the intersecting point between two BLG-like bands is broken. The dashed black lines in (e) indicate the energy levels corresponding to the Lifshitz transition points. (f) Evolution of Fermi surfaces at $\Delta = 25$ mV as a function of Fermi energy.

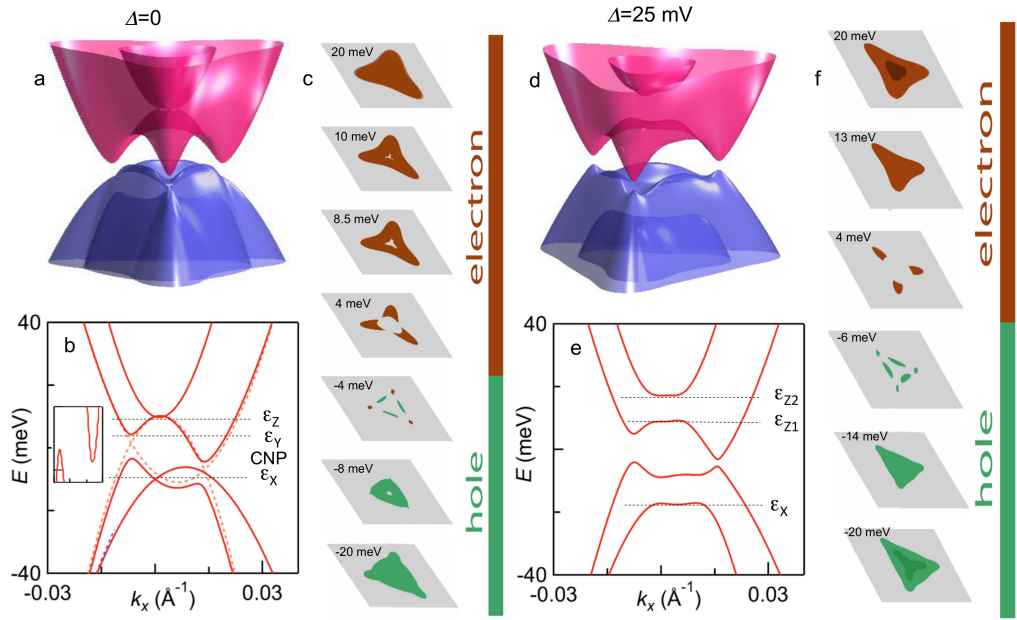

FIG. 3. (a). Experimental data map $R(n, B)$ at $D = 0$. Numbers and arrows indicate filling factors. Blue (white) regions represent the incompressible (compressible) states. (b). Calculated LL spectra for $B \leq 3T$. Numbers are LL indices. (c) High resolution data of $dR/dB$ at low magnetic field and low charge density. Dotted box indicate the crossings of triple degenerate LLs at very low field. Numbers and arrows indicate the triply degenerate LLs at filling factors $\nu = 12$ and $24$. (d). Simulation DOS $(n, B)$ for the same range of (a).

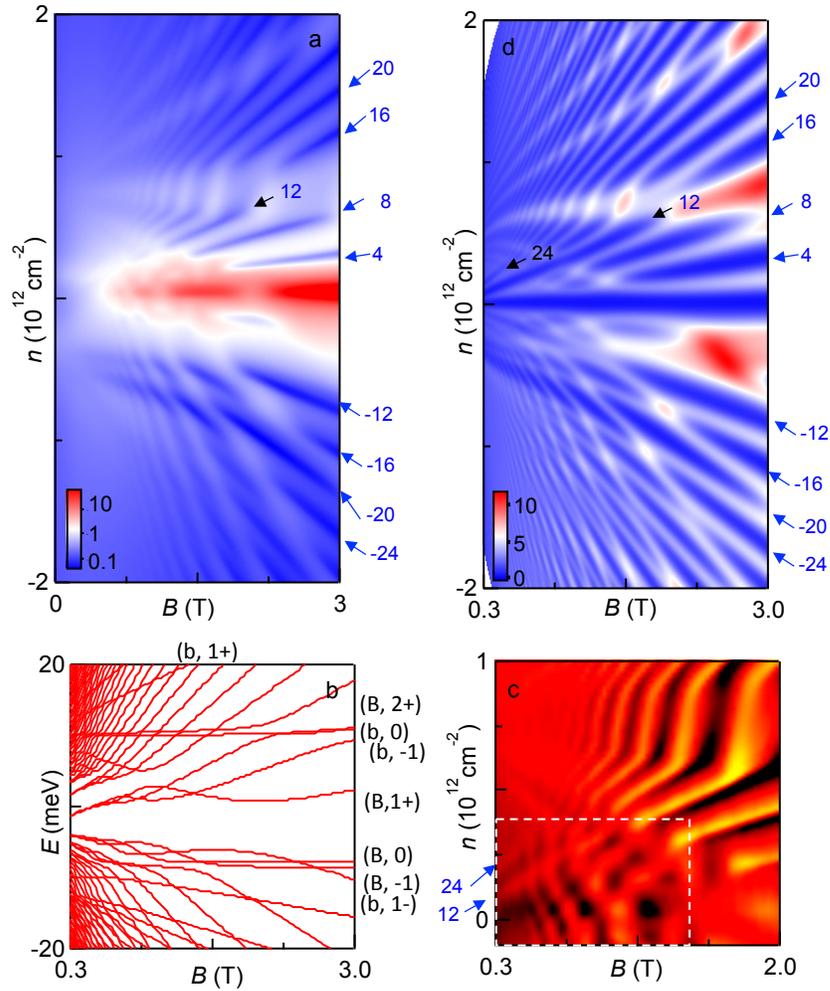

FIG. 4. (a) Experimental longitudinal resistance map $R_{xx}(\nu, D)$ at $B = 2$ T in log scale. (b-c) Calculated LL spectra and DOS, respectively, as a function of energy $E$ and $D$ at $B = 2$ T. In (b), red and blue lines stand for two valleys. In (c), the energy $E$ is converted to filling factors. Yellow starts indicate the equally spaced LL crossing points at $D = 0$.

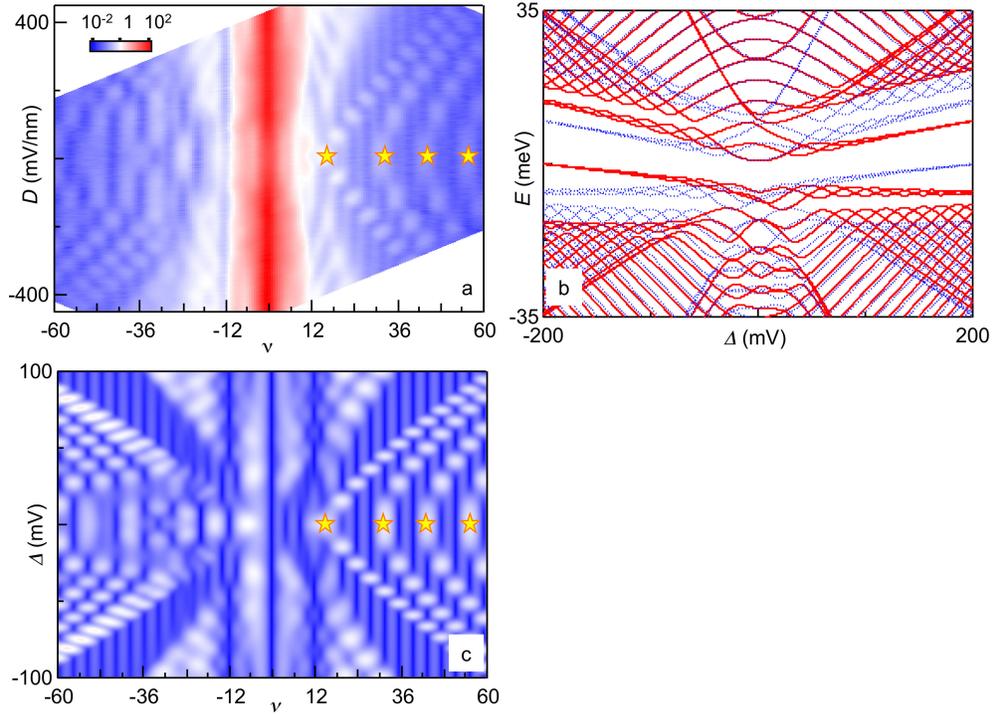